\documentclass[twocolumn]{aastex62}

\received{January 1, 2018}
\revised{January 7, 2018}
\accepted{\today}
\submitjournal{ApJL}

\shorttitle{UV-X-ray spectal index for ULXs}
\shortauthors{Sonbas, Dhuga, \& G\"o\u{g}\"u\c{s}}

\begin{document}

\title{Evidence of an X-Ray-UV Spectral Correlation in ULXs}

\email{edasonbas@gmail.com}

\author{E. Sonbas}
\affil{Adiyaman University, Department of Physics, 02040 Adiyaman, Turkey}
\affil{Department of Physics, The George Washington University, Washington, DC 20052, USA}

\author{K. S. Dhuga}
\affil{Department of Physics, The George Washington University, Washington, DC 20052, USA}

\author{E. G\"o\u{g}\"u\c{s}}
\affil{Sabanc\i~University, Orhanl\i~- Tuzla, Istanbul 34956, Turkey}



\begin{abstract}
\noindent
By comparing the ratio of flux densities in the X-ray and UV wavebands by way of the spectral optical-X-Ray index, $\alpha_{ox}$, we explore the relation between the emissions in the respective wavebands for a number of ULXs with known optical counterparts. We present a significant (anti)correlation between $\alpha_{ox}$ and the $L({2500\AA})$-UV luminosity. In comparison with low-z AGN, for which a similar correlation is observed, the  ULX $\alpha_{ox}$ indices follow a steeper slope albeit with a large uncertainty. The results are also compared with a small sample of dwarf-galaxy data consisting of a mixture of broad-line candidate AGN and 'composites'. A number of these sources follow the steeper slope of the ULX data, potentially hinting at an intrinsic similarity of these sources to ULXs. We are able to reproduce the general trend of the ULX correlation with the use of a multicolor accretion disk coupled to a hot corona of Comptonizing electrons.  
\end{abstract}



\section{Introduction} \label{sec:intro}
\noindent
Ultra-luminous X-ray sources (ULXs) have been studied for decades in the X-ray waveband ever since their discovery in the early 80's \citep{1981ApJ...246L..61L, 1989ARA&A..27...87F}. Several possibilities as to the nature of these intriguing objects continue to be discussed in the literature (see recent review by \citet{2017ARA&A..55..303K}). Current models tend to favor stellar-mass black holes (BHs) with a possible combination of effects such as relativistic beaming, and/or accretion at super-Eddington limits (\citep{2001ApJ...552L.109K, 2002ApJ...568L..97B, 2007Ap&SS.311..203R, 2015NatPh..11..551F} and references therein). Moreover, the detection of pulsations in a handful of sources \citep{2014Natur.514..202B, 2016ApJ...831L..14F, 2017Sci...355..817I, 2018arXiv181200662C}, means that at least a fraction of these sources host neutron stars, implying an overall heterogeneous underlying population as opposed to a single class of objects.\\
\\
ULXs exhibit different spectral and timing behavior in comparison with (ordinary) galactic BHs, featuring two-component X-ray spectra with soft excess and a turnover at energies near 5 keV. As well, in contrast to galactic BH binaries, ULXs tend to be more persistent rather than transient. Moreover, ULXs found in elliptical galaxies tend to be have low levels of variability than those found in star-forming galaxies, that can vary by an order of magnitude \citep{2006ApJ...653..536F}. \citet{2013MNRAS.435.1758S} extracted fractional variability and constructed variability-hardness diagrams to distinguish three main states i.e., the broadened disk (possibly dominated by a slim disk), hard ultraluminous(low-inclination i.e., more face-on), and the soft ultraluminous (high inclination i.e, more edge-on). Variabilities reaching 25-40 \%\ have been reported for these states. The three ULX spectral states do not match the states typically associated with galactic BH binaries. In addition, \citet{2009MNRAS.398.1450K} find that the `soft excess' reported for many ULXs around $\sim$ 0.2 keV does not follow the expected temperature profile of a standard disk i.e., L $\propto$ T$^4$. These considerations raise very important questions regarding the configuration/contribution of the accretion disk in ULXs.\\ 
\\
With the aid of recent HST data \citep{2013ApJS..206...14G}, significant progress has been made toward identifying optical counterparts of ULXs. Unique optical counterparts have been identified for about a dozen ULXs, with their colors and magnitudes in the optical wavebands extracted \citep{2004MNRAS.351L..83K, 2004ApJ...602..249L, 2005ApJ...620L..31K, 2012ApJ...745..123G}. These data further enable the modeling of ULXs, especially the optical emissions arising from the companions as well as contributions, direct and indirect i.e., irradiation emissions from the associated accretion disks. The availability, allbeit rather limited multiwaveband (X-ray and optical) data, have led to considerable impetus toward a re-evaluation, and possible modifications, of the standard disk model \citep{2007MNRAS.376.1407C, 2009MNRAS.392.1106G, 2009MNRAS.398.1450K, 2013AstBu..68..139V}.\\ 
\\
The overall picture that seems to be emerging regarding the disk is that of 'blobs' of matter and/or radially stratified layers of material that form an inflow toward the central BH from the companion star (see \citet{2015MNRAS.447.3243M}, and references therein). The `structured' regions can in principle serve as the source of the variability in emission observed in some of the ULX states. Further out in radius where the disk is assumed to be tapered are the regions where emissions are expected to be dominated by direct UV and optical (i.e., viscosity related heating of the disk). An additional source of UV and optical is through irradiation of the disk via the X-ray emission that originates in the interior regions of the disk and in the corona,  a hot gas, consisting primarily of energetic electrons, surrounding the central object. In this scenario the main drivers of the radiation emission are a radially stratified disk and a hot corona i.e., coupled media with varying optical density, and the expectation is a degree of correlation in the X-ray and UV bands. A somewhat different picture is presented by \citet{2018ApJ...856..128W} (and others i.e. \citet{2009MNRAS.397.1836G}) in their broadband spectral analysis of pulsar-like components in a sample of ULXs. These authors find that the majority of the emission (at least in the 0.3 - 10 keV range) is well described by a combination of two thermal components, one arising from the outer regions of the disk, and the other from supercritical flow within the spherization radius. This interpretation would seem to enhance the contribution of the disk relative to that of the corona. Regardless of the underlying mechanism, we pursue the question of whether a correlation exists between the X-ray and UV emissions, and whether, it can be understood in terms of a simple multicolor disk model.\\
\\
In this study, we explore the relation between the X-ray and optical emissions for a number of ULXs by utilizing the available X-ray and optical spectral data. We extract the X-ray and UV flux densities at 2~keV and 2500 $\AA$ respectively. From these spectral densities we compute the optical-X-ray spectral index, $\alpha_{ox}$, as defined by \citet{1979ApJ...234L...9T}.  The spectral index, as a function of luminosity, can then be compared with those of  other accreting systems, such as low-z AGN and broad-line candidate AGN and 'composites' in nearby dwarf galaxies for which similar studies \citep{2013A&A...550A..71V, 2017ApJ...836...20B} have been performed. The layout of the paper is as follows: in Section 2, we describe the selection and extraction of the available X-ray and optical data; Section 3 contains the details of the computation of the spectral index and its comparison with the results from a sample of AGN, and lastly, we conclude by summarizing our main findings in Section 4.
\section{Sample Selection and Data Reduction}
\noindent
\citet{2013ApJS..206...14G} identified 45 ULXs from a number of ULX catalogs available in the public domain for which $Chandra$ and $Hubble~Space~Telescope$ (HST) data are available. Given that our primary aim is to probe possible UV-X-ray spectral correlations, we focus our attention on those ULXs for which the optical counterpart is uniquely known. Ideally, the datasets should be based on simultaneous or quasi-simultaneous observations for each source in the X-ray and UV bands to ensure minimization of extraneous (incoherent) contributions to the emissions in the different wavebands. This would  allow a more robust probe of the intrinsic variability of each source, however, this was not the case in most of the datasets for which we could find HST observations in the UV filters i.e., F300W and F336W (Wide Field Planetary Camera 2), and F330W and F220W (ACS High Resolution Camera (ACS-HRC)). In addition, we wish to keep the potential background UV emissions from the host galaxies to a minimum, which limits us to HST data sets because of the superior spatial resolution of HST compared to the Swift(UVOT) and XMM-Newton(OM) instruments. With these criteria applied, we have a reduced data set of 9 ULXs (and 10 observations) for which appropriate HST(UV) observations are available; these are listed in Table 1. The selected sources show no evidence of X-ray pulsations, however, this does not eliminate the possibility that the sources host NSs. The UV (intrinsic) magnitudes were taken directly from \citet{2013ApJS..206...14G} and were used to obtain the UV flux density at 2500$\AA$. We note that the peaks of the UV filters are not strictly at 2500 $\AA$ and an extrapolation, assuming a powerlaw spectral model, was used to extract the appropriate flux densities (see below). This is potentially a source of dispersion in the extracted UV luminosity density.\\
\\
For the X-ray data, we downloaded the longest appropriate Chandra observation for each ULX source in our sample. The observation IDs and instrument setup for each of the X-ray and UV observations are listed in Table 1. For the Chandra data set, we used version 4.10 of CIAO (with CALDB 4.7.8) to reduce the data. Using the specextract script, source and background spectra and responses were created from each observation. The source spectra were extracted using a circular region centered on the source. Background spectra were extracted from an adjacent, source-free, region of the same size. The HEASARC/FTOOLS suite was used to further process the data; the routine grppha was used to group all spectra to include at least 5 counts per bin.
\section{Results and Discussion}
\noindent
As noted elsewhere, ULXs are often compared to galactic BH binaries but one can also attempt a comparison with AGN and examine the possibility of spectral similarities at larger scales in both luminosity and mass of the central object. This is partially justified because of recent studies \citep{2006Natur.444..730M} indicate `universal' scaling of some spectral properties of AGN and galactic BH binaries. The majority of the ULX population, presumably hosting stellar mass BHs, in principle, lies in the middle of this `universal' scaling in luminosity and mass. That being the case, one naturally wonders whether ULXs follow a similar scaling too, and if so, how best to probe that scaling?\\
\\
One possible way is suggested by the $\alpha_{ox}$ spectral index and its correlation observed in AGN. It is well established \citep{1982ApJ...262L..17A, 2003ApJ...588..119B, 2005AJ...130..1961S, 2006AJ...131..2826S} that the X-ray-to-UV ratio of AGN gives direct information on important regions of the spectral energy distribution, relating the radiative processes that operate in the accretion disk and in the corona, connecting their emissions across the UV and optical bands. The X-ray/UV ratio is defined \citep{1979ApJ...234L...9T} as the following inter-band spectral index:\\
\\
\begin{equation}
\alpha_{ox} \equiv  \frac{log(L_{X}/L_{UV})}{log(\nu_{X}/\nu_{UV})} = 0.384~log(\frac{L_{X}}{L_{UV}}) \\
\end{equation}
\\
where \\ 
\\
\begin{equation}
\nu_{X} \equiv \nu_{2keV} ~and~ \nu_{UV} \equiv \nu_{2500~\AA}.
\end{equation}
\\
The UV flux is measured at 2500$\AA$ primarily because this region is relatively free of strong emission lines and therefore represents a reliable measure of the continuum in the region. The X-ray flux is taken at 2 keV. The $\alpha_{ox}$ index is known to be strongly anti-correlated with the UV luminosity (\citet{2010A&A...512A..34L} and \citet{2013A&A...550A..71V}): see Figure 1, where we have reproduced the data from \citet{2013A&A...550A..71V} for a large sample of low-redshift AGN detected by Swift. Although the dispersion is relatively large, the correlation is significant. As far as we are aware, the possible existence of such a correlation for ULXs has not been reported before. Also displayed in Figure 1 (black points) are the results from a small sample of dwarf-galaxy data consisting of a mixture of broad-line candidate AGN and 'composites'  \citep{2017ApJ...836...20B}. Composite sources exhibit contributions to narrow-line emission both from the AGN and through episodes of star-formation. Once again, the dispersion is relatively large (as well as large uncertainties) but more interestingly, some of these sources exhibit significantly smaller values of $\alpha_{ox}$, and a different slope compared to the low-z AGN sample. \citet{2017ApJ...836...20B} noted both of these points.  In the following, we test the conjecture of an AGN-like $\alpha_{ox}$ correlation for ULXs by determining the spectral index for our sample of ULXs.\\ 
\\
The X-ray spectra were analyzed using XSPEC version 12.10.1 \citep{1996ASPC..101...17A}. The data were fitted over the 0.3 - 10.0 keV energy range: models considered include the PL (power-law) and/or thermal models, such as diskbb and bbody, along with galactic and intrinsic absorption. Galactic H-column densities were estimated using HEASARC nH tool\footnote{https://heasarc.gsfc.nasa.gov/cgi-bin/Tools/w3nh/w3nh.pl}. Following \citep{2017ApJ...836...20B}, we extracted unabsorbed fluxes in 2 - 10 keV band.  Only in one case (NGC 4395 X-1) we needed to resort to using webPIMMS to extract the flux density at 2 keV using a powerlaw with $\Gamma$ = 2. The uncertainty in using this approach can be estimated by varying $\Gamma$ in a range that is typical for BH binary systems in the relatively high luminosity regime i.e., $\sim$ 1.7 - 2.0. The 2 keV flux changes by approximately 10$\%$, which is consistent with the uncertainties determined directly from the data. The flux density at 2500$\AA$ was calculated in two steps: Initially, using the intrinsic magnitudes (for the UV filters (F336W, F330W or F300W) as given by \citet{2013ApJS..206...14G}), we determined the flux density at the appropriate midpoint wavelength of the given filter. We then extrapolated this flux density to the 2500$\AA$ point by assuming a powerlaw spectral model with an index of 1/3 (which is consistent with a standard disk). Typically the extrapolated flux densities differ from those corresponding to the filter midpoints by 15-20\%. The $\alpha_{ox}$ index was calculated using the expression given above, and is plotted (red points) as function of the 2500$\AA$  UV-luminosity density in Figure 1. Although the error bars are large, the trend of the data is clearly suggestive of a correlation. The best-fit relation for $\alpha_{ox}$ - $L_{2500\AA}$ is \\
 \\
 $\alpha_{ox}$ = (-0.311 $\pm$ 0.061)~\textit{log}~$L_{2500\AA}$ +  (6.61 $\pm$ 1.41) \\
 \\
\noindent 
with a slope of -0.311 $\pm$ 0.061, and is indicated by the solid black line. For comparison, also shown (grey points) is $\alpha_{ox}$ for low-redshift AGN sample of \citet{2013A&A...550A..71V} with a significantly shallower slope of  -0.135 $\pm$ 0.015 (solid blue line). However, with a $\sigma$ of 0.061, the best-fit ULX slope is technically within 3$\sigma$ of being consistent with the AGN data. A fit to the combined ULX/AGN data returns a slope of -0.121 $\pm$  0.015, which is very close to that given by \citet{2013A&A...550A..71V}. This is not surprising as the fit is dominated by the large and more precise AGN data.  Although an unlikely scenario, given the large $\sigma$ and the very limited dynamic range of the ULX data, one cannot strictly rule out the possibility that the two data sets are consistent. \\
\\
Furthermore, we note that a number of sources in the dwarf-galaxy sample of candidate broad-line AGN and 'composites' follow the ULX slope instead of the AGN sample of \citet{2013A&A...550A..71V}. This is illustrated by the extended dashed black line. 
This apparent consistency is intriguing and hints at an intrinsic similarity of these particular sources to ULXs. The similarity is unlikely due to the mass of the compact object since we expect the majority of ULXs to be stellar-mass BHs whereas \citet{2017ApJ...836...20B} report the mass of their sources to lie in the range 10$^{5} - 10^{6} M_\odot$. 
For the AGN sample the dissimilarity of the slope may simply be due to the fact that the AGN are sub-Eddington sources whereas the ULXs are likely to be in the super-Eddington regime. This does not necessarily imply that the few of the \citet{2017ApJ...836...20B} sources that appear to follow the ULX correlation are also high accreting systems because enhanced UV contributions as a result of nuclear star formation in these sources could produce smaller values of $\alpha_{ox}$. In fact, \citet{2017ApJ...836...20B} identify one of these sources as a star forming object and it lies on the ULX correlation.\\
\\
We also examined $\alpha_{ox}$ as a function of the $L_{2 keV}$ luminosity  but found little evidence for a correlation i.e., the slope is consistent with zero within error bars (plot not shown). Moreover, unlike the AGN data, which exhibit a reasonably tight correlation between the $L_{2 keV}$ and $L_{2500A}$ luminosities (\citet{2010A&A...512A..34L} and \citet{2013A&A...550A..71V}), the ULX data show considerable dispersion, possibly indicative of the intrinsic variability noted by \citet{2013MNRAS.435.1758S}. For example, the UV luminosity of the source Hol IX-X1 (see Table 1) varies by an order of magnitude in two observations. This not only produces dispersion in the aforementioned plot but also strongly argues for simultaneous X-ray and UV measurements if robust comparisons are to made with interband indices such as  $\alpha_{ox}$. We mention two sources for this dispersion: UV/optical contribution from the host galaxy, and perhaps more likely, the emission from the optical counterpart. The work of \citet{2005MNRAS.362...79C} seems to suggests that the stellar component dominates the disk component for low BH mass and the role reverses for high BH mass i.e., the disk component dominates for a BH mass greater than $\sim$100$M_\odot$. \\
\begin{figure}
\hspace{-1.1cm}
\centering
\includegraphics[width=8.0 cm, height=6.0cm, angle =0 ]{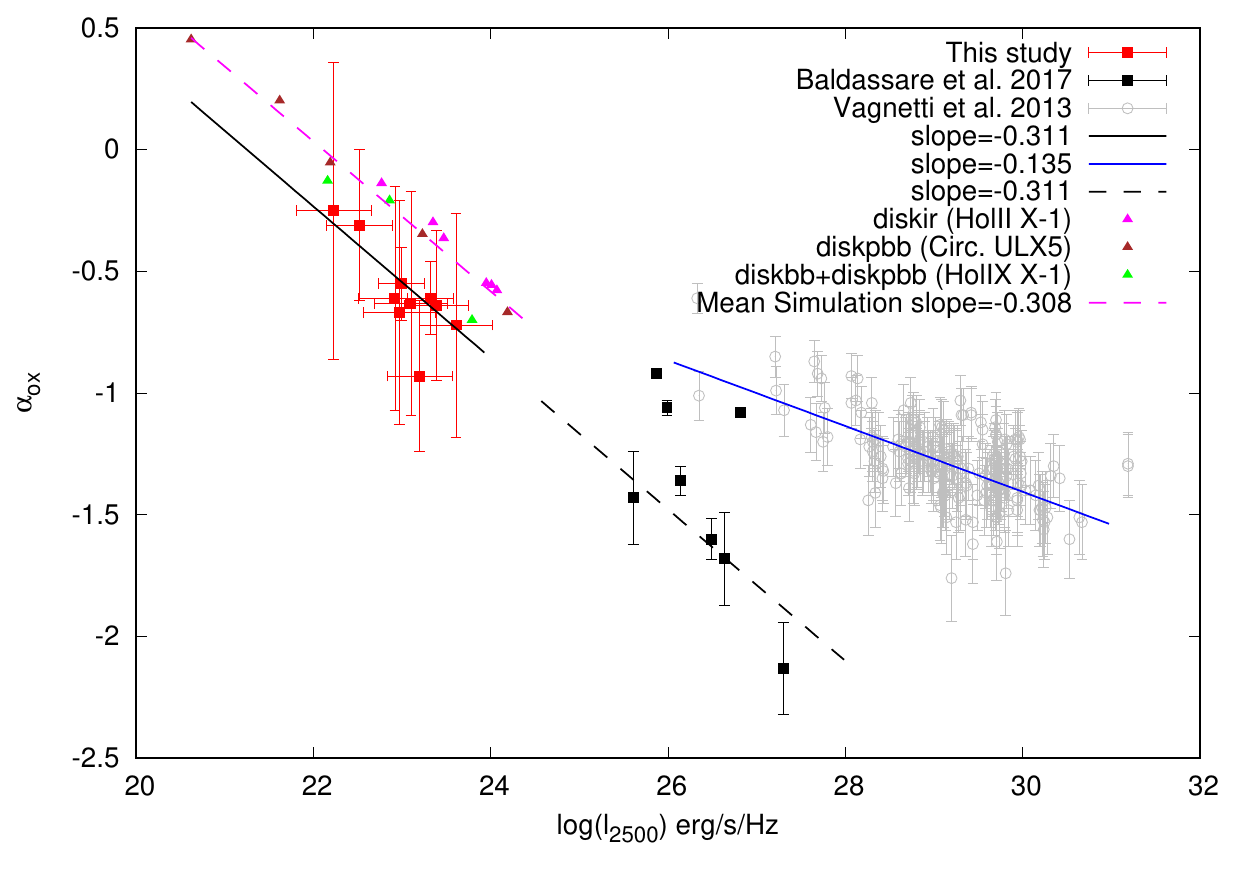}
\caption{\footnotesize{The $\alpha_{ox}$ index as a function of UV(2500$\AA$) luminosity for a) number of ULXs (red), b) a sample of low-z AGN (grey), and c) AGN and 'composites' in nearby dwarf galaxies (black points). The simulated values are shown as triangles (magenta; \textit{diskir}, brown;\textit{diskpbb}, and green;\textit{diskbb+diskpbb}). Best-fit line for the ULX data set is shown as a solid black line (extended as dashed line to include dwarf galaxy data set). Blue solid line is best-fit for the large AGN data set.}} 
\end{figure}\\
In order to explore whether the observed behavior of the spectral optical index, especially its slope, can be reproduced by a multicolor blackbody disk, we use the \textit{diskir} model \citep{2009MNRAS.392.1106G} in XSPEC to create model SEDs for ULXs. We realize this model may not be fully compatible with superaccretion flow but we note that the aim here is to extract the monochromatic luminosities at 2 keV and 2500$\AA$  respectively so that we can probe the trend of $\alpha_{ox}$ rather than to model the detailed features of any particular SED. The primary parameters in \textit{diskir} that effect the UV emission are $\it {fout}$, the fraction of bolometric flux which is thermalized in the outer disk, $\it{rout}$, the outer radius of the accretion disk, and the ratio $\it {Lc/Ld}$, which represents a measure of the Compton tail to disk luminosities. Other parameters mainly effect the X-ray emission, including $\it {kT_{in}}$ (the inner disk temperature), $\Gamma$ (powerlaw index for Comptonization), and $\it {kT_{e}}$, the temperature of the corona, and have negligible effect on the UV. For the actual parameters, we took Hol II X-1 as a representative case, for which the SED has been modeled by \citet{2013AstBu..68..139V}. We carried out a number of tests; in the first test we varied the the inner disk temperature in the range 0.1 - 0.7 keV while keeping all the other parameters fixed at those given by \citet{2013AstBu..68..139V}. In another test, we varied the ratio $\it {Lc/Ld}$ that controls the luminosity of the Compton tail relative to the disk luminosity, and finally, in the third case, we varied $\it{fout}$ in the range 0.001 - 0.1 and kept other parameters fixed at their nominal values. As expected, the results of the first test led to large increases in X-ray emission relative to UV thus producing a positive correlation for $\alpha_{ox}$ with UV luminosity contrary to that observed in the data. The second test showed very little sensitivity to the luminosities and thus the $\alpha_{ox}$ index showed no appreciable variation. The results of the third test, on the other hand, are very encouraging and are shown (as magenta points) in Figure 1, where we find a correlation that is remarkably similar to that seen in the data. \\
\\
In another round of tests, we followed \citet{2018ApJ...856..128W}, who used the \textit{diskpbb} model \citep{1994ApJ...426..308M} to demonstrate the importance of thermal components in ULX spectra; in this model the local disk temperature is represented by a powerlaw with an exponent $p$ that can be varied. They found values of $p$ in the range 0.56 - 0.71, suggesting a much flatter temperature profile compared to the standard disk ($p$ = 0.75). Using \textit{diskpbb} and varying $p$ in the range 0.5 - 0.8, we determined $\alpha_{ox}$ for Circ ULX5, one of the Walton et al (2018) sources. The results are indicated as brown triangles in Figure 1. Also shown are the results for the source Hol IX X-1 (green triangles) with parameters of \citet{2018ApJ...856..128W}. Clearly, the trend of $\alpha_{ox}$ is reproduced. In replicating the trend, both the \textit{diskir} and the \textit{diskpbb} tests strongly point to the accretion disk, and possibly its structure, as the likely cause of the observed correlation.  However, we refrain from drawing a firm conclusion as the UV/optical contribution from the optical counterpart, which could be substantial for a given combination of  stellar companion and BH mass, has not been explicitly taken into account. \\
\section{Summary and Conclusions}
\noindent
In this paper, we have extracted the $\alpha_{ox}$ spectral index for a sample of (9) ULXs for which unique optical counterparts are known. The calculation of the index requires the monochromatic luminosities at 2500 $\AA$  and 2 keV respectively. We used the UV magnitudes given by \citet{2013ApJS..206...14G} to extract the UV flux density at 2500 $\AA$ . The 2 keV X-ray flux density was determined from spectral fits to Chandra data sets for the sample ULXs. We compared our results with those of low-z AGN data of \citet{2013A&A...550A..71V} and the dwarf-galaxy sample of \citet{2017ApJ...836...20B}, comprising of candidate broad-line AGN and 'composites'. The \textit{diskir} and \textit{diskpbb} models were used to explore the coupling between the emissions in the X-ray and UV bands. We summarize our main findings as follows:
\begin{itemize}
\item We find evidence for a significant anticorrelation between the $\alpha_{ox}$ index and the UV monochromatic flux density at 2500$\AA$ for a number of ULXs 
\item The ULX correlation is similar to that observed for low-z AGN but with a significant steeper slope i.e., -0.311$\pm$0.061 compared to -0.135 $\pm$ 0.015. However, consistency between the two best fits is not strictly ruled out given the large uncertainties in the ULX data
\item The best-fit to the ULX data set, extended as a dashed line (Figure 1), is (partially) consistent with the dwarf-galaxy data set of \citet{2017ApJ...836...20B}. These data set contain a mixture of candidate AGN and 'composites'. Three (composite) sources apppear to follow the standard AGN trend line (\citet{2010A&A...512A..34L} and \citet{2013A&A...550A..71V}) while the other five sources (4 broad-line AGN candidates + a star-forming object) are more consistent with the steeper trend of the ULX data, potentially hinting at a ULX-like nature for these particular sources
\item We are able to reproduce the main trend of the observed ULX correlation with two accretion disk models: \textit{diskir} and \textit{diskpbb}. The first one takes into account the spectral emissions (including reprocessing) from a multicolor accretion disk coupled to a hot corona of Comptonizing electrons, and the second one allows the probe of advection flow through a variation in the temperature profile of the disk. We note the results reported here represent an exploratory set of calculations and a more robust investigation awaits additional multiwavelength data (ideally from synchronized observations) and requires greater scrutiny of the parameter space of \textit{diskir}, \textit{diskpbb}, and more sophisticated models (e.g., SCAD; \citet{2013AstBu..68..139V}) especially constructed for investigating superaccretion flows. A parameter of particular importance of course is the mass of the compact object as it has important implications in the search for intermediate mass BHs
\item Finally, we end on a note of caution that the stellar contribution to the UV component from the counterpart has not been taken into account and could be substantial depending on the BH mass and the nature of the companion.    
\end{itemize}
\begin{deluxetable*}{crrrrrrr}
\tablecaption{UV and X-ray properties of sources in the sample}
\tablehead{\colhead{Source Name} &\colhead{Distance} &\colhead{Chandra } & \colhead{HST Filter} & \colhead{Magnitude\tablenotemark{1}} &\colhead{f$_{2 keV}$} & \colhead{f$_{2500 \AA}$}  & \colhead{$\alpha_{ox}$}  \\
\colhead{} & \colhead{Mpc} & \colhead{Obs. ID} & \colhead{} & \colhead{} & \colhead{(erg cm$^{-2}$ s$^{-1}$ $Hz^{-1}$)} & \colhead{(erg cm$^{-2}$ s$^{-1}$ $Hz^{-1}$)} & \colhead{} }
\startdata
 NGC4190 X-1&2.80 &8212 & WFPC2/F300W &22.0 $\pm$ 4.0 & 3.95$\pm$1.55 $\times 10^{-30}$ &1.82$\pm$1.77 $\times 10^{-29}$ & -0.25$\pm$0.61 \\
 M81 X-6 & 3.40&735 &  WFPC2/F336W &21.0 $\pm$ 3.0& 1.53$\pm$0.05 $\times 10^{-30}$ & 5.95$\pm$5.58 $\times 10^{-29}$& -0.61$\pm$0.46 \\
  HOL IX X-1& 3.42& 9540 &ACS–HRC/F330W  &20.0 $\pm$ 1.0& 3.81$\pm$0.11 $\times 10^{-30}$ & 1.50$\pm$0.90 $\times 10^{-28}$& -0.61$\pm$0.15 \\
  -- & &9540 & ACS–HRC/F330W  & 22.0 $\pm$ 2.0 &3.81$\pm$0.11 $\times 10^{-30}$  & 2.39$\pm$2.00 $\times 10^{-29}$&  -0.31$\pm$0.31\\
  NGC4395 X-1  & 3.60&402 &WFPC2/F336W &21.0 $\pm$ 3.0 &1.10$\pm$0.11 $\times 10^{-30}$  & 5.95$\pm$5.58 $\times 10^{-29}$& -0.67$\pm$0.46 \\
NGC1313 X-1     & 3.70&2950 &ACS–HRC/F330W  &21.0 $\pm$ 1.0 &2.21$\pm$0.12 $\times 10^{-30}$  & 5.99$\pm$3.61 $\times 10^{-29}$&-0.55$\pm$0.15  \\
 NGC1313 X-2     & 3.70&3550 & ACS–HRC/F330W &20.0 $\pm$ 2.0&3.28$\pm$0.18 $\times 10^{-30}$  & 1.50$\pm$1.26 $\times 10^{-28}$& -0.64$\pm$0.31 \\
 NGC2403 X-1      & 4.20&2014  &ACS–HRC/F330W  &21.0 $\pm$ 3.0& 1.35$\pm$0.07 $\times 10^{-30}$ & 5.99$\pm$5.62 $\times 10^{-29}$& -0.63$\pm$0.46 \\
  M83 X-2\tablenotemark{2}      & 4.70&793 & WFPC2/F336W &21.0 $\pm$ 2.0&2.28$\pm$0.42 $\times 10^{-31}$ & 5.95$\pm$5.00 $\times 10^{-29}$& -0.93$\pm$0.31  \\
  NGC5204 X-1       &  4.80&3943 &ACS–HRC/F220W  &19.0 $\pm$ 3.0& 2.00$\pm$0.23 $\times 10^{-30}$  & 1.49$\pm$1.39 $\times 10^{-28}$& -0.72$\pm$0.46 \\
\enddata 
\label{table:nonlin} 
\tablenotetext{1}{Intrinsic magnitudes were taken from \citet{2013ApJS..206...14G}}
\tablenotemark{2}{Intrinsic magnitude not available. Plotted in Figure 1 using corrected magnitude but not used in obtaining best fit.}
\end{deluxetable*}
\acknowledgements
We acknowledge the anynomous Referee for the constructive comments that helped to improve the manuscript.
\software{ciao (4.10), XSPEC (v12.9.1; Arnaud 1996)}


\begin{thebibliography}{}

\bibitem[Arnaud(1996)]{1996ASPC..101...17A}Arnaud, K., 1996, ASPC, 101, 17

\bibitem[Avni \& Tananbaum(1982)]{1982ApJ...262L..17A}Avni, Y.; Tananbaum, H.  1982, ApJ, 262, 17 

\bibitem[Bachetti et al.(2014)]{2014Natur.514..202B}Bachetti, M.; Harrison, F. A.; Walton, D. J.; Grefenstette, B. W.; Chakrabarty, D. et al. 2014, Nat, 514, 202

\bibitem[Baldassare et al.(2017)]{2017ApJ...836...20B}Baldassare, V. F.; Reines, A. E.; Gallo, E.; Greene, J. E., 2017, ApJ, 836, 20  

\bibitem[Bechtold et al.(2003)]{2003ApJ...588..119B}Bechtold, J., Siemiginowska, A., Shields, J., et al. 2003, ApJ, 588, 119

\bibitem[Begelman(2002)]{2002ApJ...568L..97B}Begelman, M. C. 2002, ApJ, 568, L97 

\bibitem[Carpano et al.(2018)]{2018arXiv181200662C}Carpano, S.; Haberl, F.; Crowther, P.; Pollock, A., 2018arXiv181200662C

\bibitem[Copperwheat et al.(2005)]{2005MNRAS.362...79C}Copperwheat, C.; Cropper, M.; Soria, R.; Wu, K. 2005, MNRAS, 362, 79

\bibitem[Copperwheat et al.(2007)]{2007MNRAS.376.1407C}Copperwheat, C.; Cropper, M.; Soria, R.; Wu, K., 2007, MNRAS, 376. 1407 

\bibitem[Fabbiano(1989)]{1989ARA&A..27...87F} Fabbiano, G. 1989, ARA\&A, 27, 87

\bibitem[Fabrika et al.(2015)]{2015NatPh..11..551F}Fabrika, S.; Ueda, Y.; Vinokurov, A.; Sholukhova, O.; Shidatsu, M. 2015, NatPh, 11, 551

\bibitem[Feng \& Kaaret(2006)]{2006ApJ...653..536F}Feng, H.; Kaaret, P. 2006, ApJ, 653, 536 

\bibitem[Furst et al.(2016)]{2016ApJ...831L..14F}Furst, F.; Walton, D. J.; Harrison, F. A.; Stern, D.; Barret, D. et al. 2016, ApJ, 831, L14

\bibitem[Gierlinski et al.(2009)]{2009MNRAS.392.1106G}Gierliński, M.; Done, C.; Page, K. 2009, MNRAS, 392, 1106

\bibitem[Gladstone et al.(2009)]{2009MNRAS.397.1836G}Gladstone, J. C.; Roberts, T. P.; Done, C. 2009, MNRAS, 397, 1836      

\bibitem[Gladstone et al.(2013)]{2013ApJS..206...14G}Gladstone, J. C.; Copperwheat, C.; Heinke, C. O.; Roberts, T. P.; Cartwright, T. F. et al. 2013, ApJS, 206, 14

 \bibitem[Grise et al.(2012)]{2012ApJ...745..123G}Grisé, F.; Kaaret, P.; Corbel, S.; Feng, H.; Cseh, D. et al. 2012, ApJ, 745, 123  

\bibitem[Israel et al.(2017a)]{2017Sci...355..817I}Israel, G. L.; Belfiore, A.; Stella, L.; Esposito, P.; Casella, P., 2017a, Science, 355, 817

\bibitem[Kaaret et al.(2004)]{2004MNRAS.351L..83K}Kaaret, P.; Ward, M. J.; Zezas, A. 2004, MNRAS, 351, 83

\bibitem[Kaaret et al.(2017)]{2017ARA&A..55..303K}Kaaret, P.; Feng, H.; Roberts, T. P. 2017, ARA\&A, 55, 303

\bibitem[Kajava \& Poutanen(2009)]{2009MNRAS.398.1450K}Kajava, J. J. E.; Poutanen, J. 2009, MNRAS, 398, 1450  

\bibitem[King et al.(2001)]{2001ApJ...552L.109K}King, A. R.; Davies, M. B.; Ward, M. J.; Fabbiano, G.; Elvis, M., 2001, ApJ, 552, L109

\bibitem[Kuntz et al.(2005)]{2005ApJ...620L..31K}Kuntz, K. D.; Gruendl, R. A.; Chu, Y.-H.; Chen, C.-H. R.; Still, M. et al. ApJ, 620, 31

\bibitem[Liu et al.(2004)]{2004ApJ...602..249L}Liu, J.-F., Bregman, J. N., \& Seitzer, P. 2004, ApJ, 602, 249   

\bibitem[Long et al.(1981)]{1981ApJ...246L..61L}Long, K. S.; Dodorico, S.; Charles, P. A.; Dopita, M. A. 1981, ApJ, 246, 61

\bibitem[Lusso et al.(2010)]{2010A&A...512A..34L}Lusso, E.; Comastri, A.; Vignali, C. et al. 2010, A\&A, 512, 34 

\bibitem[McHardy et al.(2006)]{2006Natur.444..730M}McHardy, I. M.; Koerding, E.; Knigge, C.; Uttley, P.; Fender, R. P. 2011, MNRAS, 417, 464  

\bibitem[Middleton et al.(2015)]{2015MNRAS.447.3243M}Middleton, M. J.; Heil, L.; Pintore, F.; Walton, D. J.; Roberts, T. P. 2015, MNRAS, 447, 3243

\bibitem[Mineshige et al.(1994)]{1994ApJ...426..308M}Mineshige, S.; Hirano, A.; Kitamoto, S.; Yamada, T. T.; Fukue, J. 1994, ApJ, 426, 308  

\bibitem[Roberts(2007)]{2007Ap&SS.311..203R}Roberts, T. P. 2007, Ap\&SS, 311, 203

\bibitem[Steffen et al.(2006)]{2006AJ...131..2826S}Steffen, A. T., Strateva, I., Brandt, W. N., et al. 2006, AJ, 131, 2826

\bibitem[Strateva et al.(2005)]{2005AJ...130..1961S}Strateva, I. V., Brandt, W. N., Schneider, D. P., Vanden Berk, D. G., \& Vignali, C. 2005, AJ, 130, 387

\bibitem[Sutton et al.(2013)]{2013MNRAS.435.1758S}Sutton, A. D.; Roberts, T. P.; Middleton, M. J., 2013, MNRAS, 435, 1758

\bibitem[Tananbaum et al.(1979)]{1979ApJ...234L...9T}Tananbaum, H., Avni, Y., Branduardi, G., Elvis, M.; Fabbiano, G. et al. 1979, ApJL, 234, L9

\bibitem[Tao et al.(2012)]{2012ApJ...750..110T}Tao, L.; Kaaret, P.; Feng, H.; Grise, F. 2012, ApJ, 750, 110 

\bibitem[Vagnetti et al.(2013)]{2013A&A...550A..71V}Vagnetti, F.; Antonucci, M.; Trevese, D. 2013, A\&A, 550, 71 

\bibitem[Vinokurov et al.(2013)]{2013AstBu..68..139V}Vinokurov, A.; Fabrika, S.; Atapin, K., 2013, AstBu, 68, 139 

\bibitem[Walton et al.(2018)]{2018ApJ...856..128W}Walton, D. J.; Fürst, F.; Heida, M.; Harrison, F. A.; Barret, D. et al. 2018, ApJ, 856, 128  

\end{thebibliography}
\end{document}